\documentclass[aps,twocolumn,prd,superscriptaddress,showpacs]{revtex4-1}

\usepackage{graphicx,color}
\usepackage{booktabs}
\usepackage{caption}
\usepackage[normalem]{ulem}

\usepackage{amsmath}
\usepackage{amssymb}

\usepackage[utf8]{inputenc}

\usepackage{mathtools}
\usepackage{amsfonts}
\usepackage{mathrsfs}
\usepackage{bbm}

\usepackage{graphicx}
\usepackage{color}
\usepackage{array}

\usepackage{placeins}
\usepackage{booktabs}
\usepackage{makecell}
\usepackage{subcaption}
\usepackage{floatrow}

\usepackage{xspace}
\usepackage{hyperref}

\usepackage{xifthen}

\def\0#1#2{\frac{#1}{#2}}

\def\eq#1{\eqref{#1}}

\def\Fig#1{Fig.~\ref{#1}}

\newcommand{\momarg}[1]{\ifthenelse{\isempty{#1}}{}{\left( #1 \right)}}

\def\dirac0{{\partial\hspace{.00cm}\llap{/}}\,}

\newcommand{\gettitle}{Locating the freeze-out curve in heavy-ion collisions}

\hypersetup{
	pdftitle={\gettitle},
	pdfauthor={Bluhm, Nahrgang, Pawlowski},
	pdfkeywords={Transport} {Low energy effective model} {Thermalisation} {Chemical freeze-out},
	bookmarksopen=true,
	bookmarksopenlevel=2,
	bookmarksnumbered=true
}

\begin{document}

\title{\gettitle}

\author{Marcus Bluhm}
\affiliation{SUBATECH UMR 6457 (IMT Atlantique, Universit\'e de Nantes, \\IN2P3/CNRS), 4 rue Alfred Kastler, 44307 Nantes, France
}
\affiliation{ExtreMe Matter Institute EMMI,
	GSI, Planckstr. 1,
	64291 Darmstadt, Germany
}

\author{Marlene Nahrgang}
\affiliation{SUBATECH UMR 6457 (IMT Atlantique, Universit\'e de Nantes, \\IN2P3/CNRS), 4 rue Alfred Kastler, 44307 Nantes, France
}
\affiliation{ExtreMe Matter Institute EMMI,
	GSI, Planckstr. 1,
	64291 Darmstadt, Germany
}

\author{Jan M. Pawlowski}
\affiliation{ExtreMe Matter Institute EMMI,
	GSI, Planckstr. 1,
	64291 Darmstadt, Germany
}
\affiliation{Institut f\"ur Theoretische Physik,
	Universit\"at Heidelberg, Philosophenweg 16,
	69120 Heidelberg, Germany
}

\begin{abstract}
Based on transport equations we argue that the chiral dynamics in heavy-ion collisions at high collision energies effectively decouples from the thermal physics of the fireball. With full decoupling at LHC energies the chiral condensate relaxes to its vacuum expectation value on a much shorter time scale than the typical evolution time of the fluid dynamical fields and their fluctuations. In particular, the net-baryon density remains coupled to the bulk evolution at all collision energies. As the mass scales of the hadrons are controlled by the chiral condensate, it is reasonable to employ vacuum masses in the statistical description of the hadron production at the chemical freeze-out for high collision energies. We predict that at lower collision energies the coupling of the chiral condensate to the thermal medium gradually increases with consequences for the related hadronic masses. A new estimate for the location of the freeze-out curve takes these effects into account. 
\end{abstract}



\maketitle

\section{Introduction}

Strongly interacting matter undergoes a phase transition or crossover from weakly interacting 
quarks and gluons at large temperatures $T$ and net-baryon densities $n$ to weakly interacting hadrons at low $T$ and $n$. The transition took place, for example, in the limit $n\simeq 0$ shortly after the big bang. This highly interesting physics of matter under extreme conditions can also be explored experimentally in high-energy nuclear collisions~\cite{Jacak:2012dx}. It includes the phenomena of chiral symmetry breaking, confinement, as well as thermalisation and chemical freeze-out through the crossover with the associated change of the dynamical degrees of freedom. 

First-principle studies of QCD thermodynamics, both on the lattice and with functional methods, show a chiral crossover for physical quark masses at vanishing or small densities. At vanishing baryon chemical potential $\mu_B$, or net-baryon density $n$, one obtains a pseudocritical temperature $T_c\approx 156$\,MeV. For baryon chemical potentials $\mu_B/T\lesssim 3$ the $\mu_B$-dependence of the chiral crossover temperature is well approximated by a quadratic function in $\mu_B$ with a small negative curvature. For more details see e.g.~\cite{Aoki:2006we, Fu:2019hdw, Fischer:2014ata, Bellwied:2015rza, Bonati:2018nut, Bazavov:2018mes, Isserstedt:2019pgx, Gao:2020qsj}, for recent overviews see~\cite{Fischer:2018sdj, DElia:2018fjp, Philipsen:2019rjq}. In the same region of small $\mu_B$ a critical end point (CEP) is excluded. At larger densities, the QCD phase structure is uncharted and is subject of current research. While lattice QCD techniques are obstructed by the sign problem, functional approaches do not suffer from conceptual problems and give access to this regime. However, for quantitative predictions about the existence and location of a CEP and/or inhomogeneous and mixed phases, the computations still need to be improved significantly.

Experimentally, the final hadronic particle multiplicities and their event-by-event distributions are accessible. These can remarkably well be described within statistical hadronization model (SHM) approaches, cf.~e.g.~\cite{Andronic:2017pug} and references therein, in which the particle production is assumed to follow thermal distributions. Phenomenologically, the particle composition of the hadronic medium is frozen at chemical freeze-out, an instance during the evolution of a heavy-ion collision when particle number changing processes cease to be effective. For collision energies $\sqrt{s_{\rm NN}}$ ranging from the HADES experiment at GSI to the LHC experiments at CERN the freeze-out temperature $T_{\rm fo}$ and chemical potential $\mu_{B,{\rm fo}}$ are obtained from thermal fits to the measured particle yields and their ratios~\cite{Eiseman:1992mh,Abbott:1994np,Ahle:1999in,Ouerdane:2002gm,Antinori:2004ee,Alt:2004wc,Adams:2005dq,Agakishiev:2009am,Abelev:2013vea,ABELEV:2013zaa,Adamczyk:2017iwn,Adamczewski-Musch:2018xwg}. At low $n$, the freeze-out conditions are intrinsically linked to the phase transition~\cite{BraunMunzinger:2003zz}. The particle distributions require information on the individual hadronic species, in particular their masses. The thermal fits work well using vacuum properties of hadrons as input, see e.g.~\cite{Andronic:2017pug,BraunMunzinger:2003zz,Florkowski:1999pz,Michalec:2001qf, Renk:2002sz,Zschiesche:2002zr, Broniowski:2003ax}. It is a longstanding question, how this can be reconciled with the equilibrium in-medium properties of QCD: at freeze-out temperatures and densities the chiral condensate, e.g.~\cite{Aoki:2006we, Fu:2019hdw, Fischer:2014ata, Bellwied:2015rza, Bonati:2018nut, Bazavov:2018mes}, and more importantly the hadron properties, e.g.~\cite{Aarts:2017rrl,Aarts:2017iai, Aarts:2018glk, Bazavov:2019www, Aarts:2019hrg}, are not close to their vacuum values. This is, in particular, true for small net-baryon densities as reached at the LHC. 

In the present work we argue that at high collision energies (small net-baryon density) and large system size, the chiral condensate decouples from the thermal distributions of the underlying medium and relaxes exponentially fast to its vacuum expectation value. Going to lower collision energies (larger net-baryon density) and/or smaller system size, the coupling to the medium is increasing and, consequently, the dynamics of the chiral condensate follows that of the medium. This scenario explains the currently observed increase of the empirically determined freeze-out temperature at STAR $\sqrt{s_{\rm NN}}=200$~GeV compared to the value at LHC energies~\cite{Andronic:2017pug} as an indication for the increasing importance of thermal effects for the dynamics of the chiral condensate. Based on the natural assumption that the freeze-out temperature decreases with the density, we arrive at a novel estimate for the freeze-out curve in heavy-ion collisions. 

\section{Chiral dynamics}\label{sec:section1}

In a heavy-ion collision the system crosses the chiral and confinement-deconfinement phase boundary leading to the formation of hadrons. Dynamical chiral symmetry breaking leads to the build-up of the hadron masses, in particular that of the proton. The dynamical build-up of the mass can be monitored by the chiral condensate $\int_V q\bar q$. It is related to the expectation value of the scalar resonance $\sigma(T,\mu_B)$, for more details in a QCD approach see~\cite{Fu:2019hdw}. 

If the chiral dynamics is fully coupled to the thermal medium, the hadron masses depend on $T$ and $\mu_B$. This implies that for fully coupled chiral dynamics the in-medium hadron masses can be significantly smaller than the vacuum masses. In particular this is true for small or vanishing density (LHC energies). Here, the freeze-out temperature is approximately the chiral crossover temperature, $T_{\textrm{fo}}\approx 156$\,MeV, see~\cite{BraunMunzinger:2003zz}. 
Based on~\cite{Fu:2019hdw} we find $\sigma(T_{\textrm{fo}},0)\approx 1/2\,\bar\sigma_0$ with the vacuum value $\bar\sigma_0 = \sigma(0,0)$ and, hence, significantly reduced hadron masses. 
We note that qualitatively similar conclusions can be drawn from recent lattice QCD studies~\cite{Aarts:2017rrl,Aarts:2017iai,Aarts:2018glk, Bazavov:2019www, Aarts:2019hrg} which indicate a significant temperature dependence of some hadron masses at $\mu_B=0$. In particular negative parity states show a strong mass reduction near $T_c$. 

In conclusion, chiral dynamics with full coupling to the fireball is in contradiction with a SHM approach using vacuum masses. However, in recent works on the chiral dynamics near the phase transition~\cite{Nahrgang2011mg,Bluhm:2018qkf} it was shown that in a more realistic scenario $\sigma$ (and other order parameter fields such as the conserved net-baryon density~\cite{Nahrgang:2018afz,Bluhm:2020mpc}) can deviate significantly from their in-medium equilibrium values.

In the present work we are most interested in the stage of the collision which hosts the process of dynamical chiral symmetry breaking and the chemical freeze-out. While the system is not necessarily in equilibrium at chemical freeze-out, this regime is well described by transport approaches. Hence the equations of motion for the field $\phi=(\sigma,\,n)$ can be obtained in the fluid dynamical limit as~\cite{Landau:1980mil,Son:2004iv,Fujii:2004jt}, 
\begin{align}
 \partial_t\sigma &= D_{\sigma\sigma}\frac{\delta F}{\delta\sigma}+D_{n\sigma}\nabla^2\frac{\delta F}{\delta n}+\xi_\sigma \,,
 \nonumber \\[1ex]
 \partial_tn &= D_{nn}\nabla^2\frac{\delta F}{\delta n}+D_{n\sigma}\nabla^2\frac{\delta F}{\delta\sigma}+\xi_n\, ,
\label{eq:evolutioneqs}
\end{align}
where $\xi_\sigma$ and $\xi_n$ are white noise fields and $F$ is the free energy functional. 
The noise field correlators are determined by the fluctuation-dissipation theorem with the requirement, that the equilibrium distribution for $\sigma$ and $n$ is given by the standard statistical expression. 
The (symmetric) Onsager coefficients $D_{\phi_i \phi_j}$ with $i,j=1,2$ determine the relaxation of the corresponding fields driven by the restoring forces $\partial F/\partial\phi_j$. 
For modes with small momentum $\mathbf{q}$, the evolution equation of $\sigma$ at leading order is proportional to $\mathbf{q}^0$ and the Onsager coefficient is simply given by the strength of relaxation $D_{\sigma\sigma}=\Gamma$. Due to net-baryon number conservation the evolution equation of $n$ at leading order is proportional to $\mathbf{q}^2$. At this order the kinetic terms in $F$ do not contribute but the evolutions of $\sigma$ and $n$ are still coupled. 

For the scenario considered here the chiefly important dynamics is that of the homogeneous mean field part $\bar\sigma$ with the mean field $\phi$, 
\begin{align} 
\phi= \bar \phi +\Delta\phi\,,\qquad \textrm{with}\qquad \frac{1}{V} 
\int_V \Delta\phi=0\,.
\end{align}
The analysis of the dynamics of $\bar{\sigma}$ alone suffices to constrain the location of the freeze-out curve significantly. A more quantitative discussion including the fully coupled dynamics of the complete system will be reported elsewhere. 

The mean field equation for $\partial_t \bar{\sigma}$ is obtained from \eq{eq:evolutioneqs} by a volume average which eliminates terms linear in the fluctuations $\Delta \phi$. This leads us to 
\begin{subequations}\label{eq:partialtbarsigma}
\begin{align}
 \partial_t \bar{\sigma} = -\Gamma \left[c_{\sigma}+\frac{\partial 
 	 F_{\rm vac}}{\partial \bar\sigma} + \alpha\bar{n}^2\bar{\sigma}+
  \Delta_{\rm fluc}\right]\,. 
\label{eq:partialtbarsigmaEqA}
\end{align}
In~\eq{eq:partialtbarsigmaEqA}, 
$c_{\sigma}$ is an explicit symmetry breaking term independent of the medium 
and $F_{\rm vac}$ is related to 
the vacuum potential for $\bar\sigma$. 
In addition, 
we find a term which is proportional to the square of the homogeneous part of the net-baryon density $\bar n$. The last term, $\Delta_\textrm{fluc}$, comprises the 
second order correlations of the fluctuations $\langle\Delta\phi_j\,\Delta\phi_k\rangle$ of the field 
$\phi=(\sigma,n)$ introduced above~\eq{eq:evolutioneqs}. 
Both the derivative $\partial F_{\rm vac}/\partial \bar\sigma$ and $\Delta_\textrm{fluc}$ are proportional to $\bar\sigma$ for small $\bar\phi$. 
Moreover, in equilibrium $\Delta_\textrm{fluc}$ contains the information about the thermal distributions and is proportional to $T/V$, see e.g.~\cite{Landau:1980mil}, 
\begin{align} \label{eq:Vdecay} 
\Delta_\textrm{fluc} \propto T/V\,.
\end{align} 
\end{subequations}
This decay with the volume is of relevance for our subsequent discussion.

Given its pivotal importance for the study of the freeze-out curve we highlight the most relevant aspects of the scenario that follows from the equations above: 
most importantly, the dynamics of $\bar{\sigma}$ is given by the leading contribution 
in the limit of small $\mathbf{q}$. Accordingly, 
the relaxation of $\bar\sigma$, and hence that of the chiral condensate,  
happens exponentially fast - quite in contrast to the fluid dynamical fields, such as $n$, which contain the thermal information of the medium. The relaxation time $\tau_\textrm{relax}$ of $\bar\sigma$ 
is much smaller than any other time scale in the evolution. In particular, it is smaller than the equilibration time for the net-baryon density $\tau_\textrm{eq}$ and the freeze-out time $\tau_\textrm{fo}$. The last important time scale is that of the medium equilibration of the chiral condensate, which is determined by the term proportional to $\bar n^2$ and the volume-dependent term $\Delta_\textrm{fluc}$ in \eq{eq:partialtbarsigma}. 

This leads us to two extreme scenarios for infinite and vanishing freeze-out volume $V$. For $V\to 0$, the mean field $\bar{\sigma}$ follows the evolution of the net-baryon density. In contrast, for $V\to\infty$ it relaxes rapidly to its vacuum value and equilibrates at infinite time. In short, for small volumes $\bar{\sigma}$ takes its equilibrium value at chemical freeze-out, for large volumes it takes its vacuum value. In the following, we discuss the physical consequences of~\eq{eq:partialtbarsigma} on the hadron masses near chemical freeze-out and the determination of freeze-out conditions in more detail. 

\section{Hadron masses at freeze-out}\label{sec:section2}

We have shown that the mean field $\bar\sigma$ of the scalar resonance relaxes exponentially fast. Moreover, its dynamics is coupled to the medium proportional to $\bar n$ and the 4d-volume via the ratio $T/V$, cf.~\eq{eq:partialtbarsigma}. 
For LHC collision energies the net-baryon density $\bar n$ is very small and the fluctuation contributions are suppressed by $T/V$ at chemical freeze-out, see Tab.~\ref{tab:FOVolume}. 
As a result, the evolution of $\bar\sigma$ at LHC energies effectively decouples from the medium, and $\bar\sigma$ relaxes via the leading order terms exponentially fast to the vacuum expectation value $\bar\sigma_0$, as do the related hadron masses. 

Accordingly, while the medium is still expanding and cooling through the chemical freeze-out, the chiral condensate and the related hadronic masses have relaxed to their vacuum values, and not to their thermal counterparts which depend on the temperature of the fireball. 
This is evident e.g.~for the proton, as its mass is proportional to the chiral condensate. The masses of pions and kaons, the pseudo-Goldstone bosons, are not directly proportional to the chiral condensate, but they are still related to its value: the presence or absence of medium effects in the chiral condensate is in one-to-one correspondence with the presence or absence of medium effects for all hadron masses, and in particular for pions and kaons.

\begin{table}[t]
	\centering
	\begin{tabular}[t]{|lr|c|c|c|}
		\hline
		\rule{0pt}{3ex}
		$\sqrt{s_{\textrm{NN}}}$& [GeV]& $17.3$ & $200$ & $2760$ \\
		\hline
		\rule{0pt}{3ex}
		$F$ && $4.1$ & $6.4$ & $10.2$ \\
		\hline
		\rule{0pt}{3ex}
		$dV/dy$& [fm${}^3$]& $1240$ & $2100$ & $5280$ \\
		\hline
		\rule{0pt}{3ex}
		$V$ &[fm${}^3$]& $5084$ & $13440$ & $53856$ \\
		\hline
		\rule{0pt}{3ex}
		$V/V_{\textrm{\,SPS}}$ && $1$ & $2.64$ & $10.59$ \\
		\hline
		\rule{0pt}{3ex}
		$T_{\textrm{fo}}$ & [MeV]& $154\pm 3.1$ & $162\pm 3.5$ & $156.5\pm 1.9$ \\
		\hline
		\rule{0pt}{3ex}
		$\mu_{B,\textrm{fo}}$ & [MeV]& $230^{+15}_{-10}$ & $24^{+3}_{-4}$ & $0.7\pm 3.8$ \\
		\hline
	\end{tabular}
	\caption[]{\label{tab:FOVolume} Estimates for the dependence of the fireball volume $V$ on the collision energy $\sqrt{s_{\textrm{NN}}}$ at chemical freeze-out, from SPS to top-RHIC to LHC. The full phase-space factors $F$ originate from the interpolation of measured $dN_{\textrm{ch}}/dy$ for one unit of rapidity around $y=0$~\cite{Adam:2016ddh}. These factors are applied to the freeze-out volumes $(dV/dy)$  deduced for one unit of rapidity~\cite{Andronic:2014zha}. The corresponding freeze-out parameters $T_{\textrm{fo}}$ and $\mu_{B,\textrm{fo}}$ from~\cite{Andronic:2017pug} are also listed.}
\end{table}

This explains why particle multiplicities at LHC energies are well described by a thermal model based on vacuum hadronic masses. Indeed, a posteriori, it provides a justification for such a procedure. 
In contrast, the decoupling of the chiral dynamics from the medium is less effective at lower collision energies. The evolution of $\bar{\sigma}$ is gradually more affected by the medium with decreasing $\sqrt{s_{\textrm{NN}}}$. Firstly, the term proportional to $\bar{n}$ becomes more relevant.  Secondly, the fluctuation terms are less suppressed for the smaller fireball volumes at the lower energies, see Tab.~\ref{tab:FOVolume}. 
While the leading order terms, in the absence of a coupling to the medium, force a quick relaxation of $\bar\sigma$ to the vacuum expectation value, the increasing coupling to the medium with lower collision energies drives $\bar\sigma$ towards its in-medium equilibrium value $\bar{\sigma}_{\rm eq}(T,\mu_B)$. As a consequence of these two competing tendencies we have 
\begin{align}\label{eq:Constraintsig}
\bar{\sigma}_{\rm eq}\leq\bar{\sigma}\leq{\bar\sigma_0}\,. 
\end{align}
For small values of $\mu_B$ (large $\sqrt{s_{\textrm{NN}}}$) the departure of $\bar\sigma$ from $\bar\sigma_0$ is given by 
\begin{equation}
 \bar{\sigma}(\mu_B) = \left(1-\kappa_\sigma\left(\frac{\mu_B}{\bar\sigma_0}\right)^2+\dots\right) \bar{\sigma}_0 \,,
\label{eq:redmasses}
\end{equation}
where the value for $\kappa_\sigma$ must be determined by imposing a physical condition, see below. Accordingly, thermal effects start to impact the related hadronic masses at the chemical freeze-out with increasing $\mu_B$ (decreasing $\sqrt{s_{\textrm{NN}}}$). 
For collision energies much lower than at the LHC, $\bar{\sigma}_{\rm eq}(T_{\rm fo},\mu_{B,{\rm fo}})$ approaches the vacuum expectation value. This is because for the smaller freeze-out temperatures at larger densities deviations from the chiral crossover temperature become more significant~\cite{Fu:2019hdw,Floerchinger:2012xd}. Moreover, the crossover gets steeper with increasing density and thus $\bar{\sigma}_{\rm eq}$ at chemical freeze-out tends towards its value at vanishing $T$. 

\begin{figure}[t]
\includegraphics[width=0.95\textwidth]{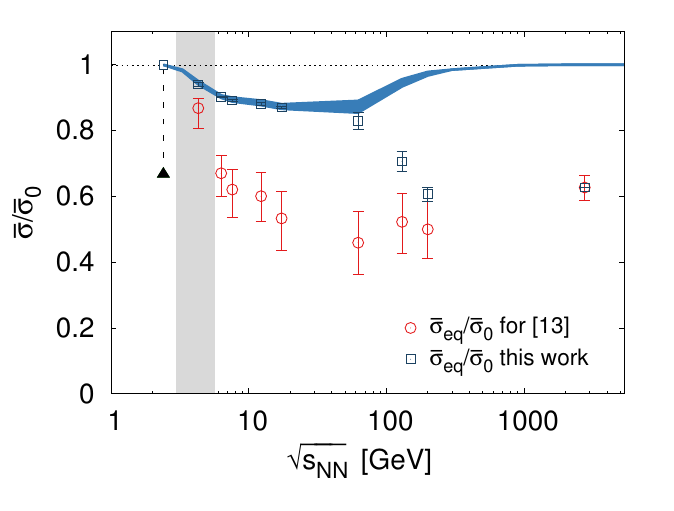}
\caption{Scaled homogeneous mean field $\bar\sigma/\bar\sigma_0$ as a function of collision energy $\sqrt{s_{\textrm{NN}}}$ at chemical freeze-out (blue filled band). In addition, the equilibrium values $\bar\sigma_{\textrm{eq}}/\bar\sigma_0$ at the freeze-out parameters from~\cite{Andronic:2017pug} (red open circles) are contrasted with those at the freeze-out conditions determined in this work (blue open squares), see \Fig{fig:PDiaChiralFOpointsMod}. The gray shaded region indicates current estimates for the location of the CEP based on functional methods, see~\cite{Fu:2019hdw} and references therein. The triangle at the lowest $\sqrt{s_{\textrm{NN}}}$ highlights our expectations for a mixed phase.}
\label{fig:ChiralCondensateAtFO2N}
\end{figure}

In \Fig{fig:ChiralCondensateAtFO2N} we illustrate the collision energy dependence of $\bar\sigma$ at chemical freeze-out as we have just discussed from the point of view of the chiral dynamics (blue filled band). At LHC  energies we find $\bar\sigma\simeq\bar\sigma_0$ due to the decoupling of $\bar\sigma$  from the medium. With increasing $\mu_B$ (decreasing $\sqrt{s_{\textrm{NN}}}$) this curve first decreases as the coupling to the medium becomes stronger. When the coupling to the medium is complete and $\bar\sigma$ relaxes to its in-medium equilibrium value $\bar\sigma_{\rm eq}$ the curve increases again by following the trend of $\bar\sigma_{\rm eq}$. In this energy range, the hadron masses at chemical freeze-out follow systematically the behavior of the dynamical chiral condensate (or $\bar\sigma$) as depicted by the blue band: for hadrons which obtain the largest part of their masses at the chiral crossover transition, we expect that at LHC energies the relevant masses at chemical freeze-out are the vacuum masses, while at lower collision energies thermal effects on the hadron masses become more important.

Keeping in mind the theoretical limitations of current calculations at larger $\mu_B$ we show in Fig.~\ref{fig:ChiralCondensateAtFO2N} a gray band in $\sqrt{s_{\rm NN}}$ which indicates the combined range of CEP locations from functional methods. Beyond this band, and in particular for $\sqrt{s_{\rm NN}}=2.4$~GeV at HADES, we can expect an onset of new physics. Still, we have indicated the equilibrium expectation $\bar\sigma_{\rm eq}$ at $\sqrt{s_{\rm NN}}=2.4$~GeV in an extrapolation of the hadronic gas phase with a blue open square. Another estimate is deduced from the low-$T$ limit: at very low temperatures and densities larger than the onset density of the liquid gas transition we arrive at a nuclear liquid. There we expect a significant drop to $\bar\sigma_{\rm eq}\approx 2/3\,\bar\sigma_0$, cf.~e.g.~\cite{Floerchinger:2012xd}. This value is shown by the triangle in~\Fig{fig:ChiralCondensateAtFO2N}. 

Based on~\cite{Fu:2019hdw} we show in \Fig{fig:ChiralCondensateAtFO2N} also the in-medium equilibrium values $\bar\sigma_{\textrm{eq}}$ at the freeze-out parameters from~\cite{Andronic:2017pug} (red open circles). For all collision energies except $\sqrt{s_{\textrm{NN}}}=4.3$~GeV we find that $\bar\sigma_{\textrm{eq}}$ is significantly smaller than $\bar\sigma_0$. These values are contrasted with the values for $\bar\sigma_{\textrm{eq}}$ at the freeze-out conditions determined in this work (blue open squares), see \Fig{fig:PDiaChiralFOpointsMod}, taking the dynamics of the chiral condensate into account.

\section{Locating the freeze-out curve}\label{sec:section3}

The in-medium equilibrium values of the homogeneous mean field in~\Fig{fig:ChiralCondensateAtFO2N} (red open circles) are obtained for the freeze-out parameters~\cite{Andronic:2017pug}. These base on fits using vacuum hadronic masses. As we have argued this is well justified at LHC energies. However, as $\mu_B$ starts to increase the dynamical chiral condensate decreases. Therefore the masses of hadrons, which obtain their mass at the chiral crossover transition, are smaller than in vacuum. It is clear that for obtaining the same measured particle multiplicities within a thermal fit the freeze-out temperature must drop accordingly. This can be seen from the Boltzmann factors $\exp{(-m_i/T)}$ in the thermal distributions. To illustrate the effect more quantitatively, we keep $\mu_{B,{\rm fo}}$ fixed for simplicity in the following. 

According to the behavior of the dynamical chiral condensate, cf.~\Fig{fig:ChiralCondensateAtFO2N}, in-medium effects already play a role at the highest STAR collision energy, $\sqrt{s_{\rm NN}}=200$~GeV. Thus, performing a thermal fit with vacuum masses at this energy leads to a freeze-out temperature which is over-estimated and therefore larger than at LHC. Since we know from lattice QCD calculations that the pseudocritical temperature $T_c$ decreases as a function of $\mu_B$ (and at this energy $T_{\rm fo}$ is intrinsically linked to $T_c$), we can determine the coefficient $\kappa_\sigma$ in the expansion around $\mu_B=0$ in \eq{eq:redmasses} by requiring that the slope of the freeze-out curve at $\mu_B\simeq 0$ is vanishingly small. In practice, we require that the freeze-out temperature at the highest STAR collision energy should be equal or less than the freeze-out temperature at LHC: $T_{\rm fo}^{\rm STAR} \leq T_{\rm fo}^{\rm LHC}$. 

\begin{figure}[tb]
\includegraphics[width=0.95\textwidth]{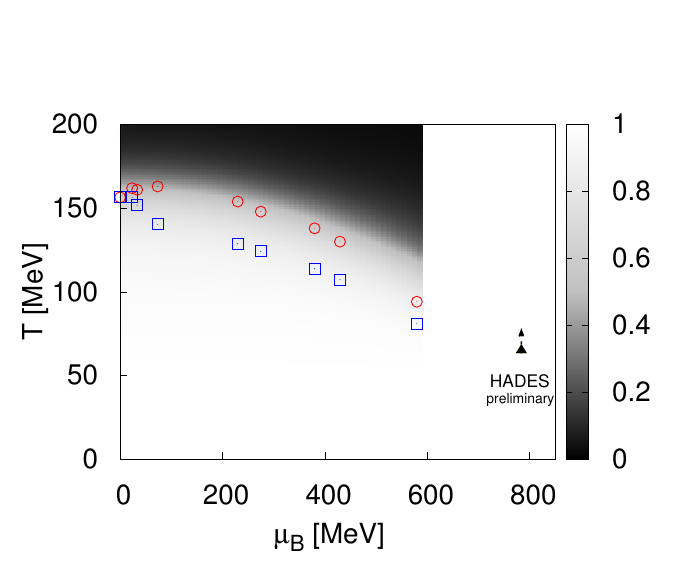}
\caption{Chemical freeze-out parameters from~\cite{Andronic:2017pug} (red open circles) in comparison with the freeze-out conditions determined in this work (blue open squares). The values of the scaled homogeneous mean field in equilibrium $\bar\sigma_{\textrm{eq}}/\bar\sigma_0$ (right scale) in the QCD phase diagram based on~\cite{Fu:2019hdw} are also shown. In addition, the triangle at very large $\mu_B$ indicates current estimates~\cite{HADESFO1,HADESFO2} for the freeze-out conditions at the HADES collision energy.}
\label{fig:PDiaChiralFOpointsMod}
\end{figure}

For our estimates we focus on the proton multiplicities only. The proton mass $m_p$ is to a large degree determined by the value of the chiral condensate. It is a good average representative for the in-medium effects on all hadrons: on the one hand, the pion and the kaon (as pseudo-Goldstone bosons) and the higher mass resonances are less affected. On the other hand, the masses of negative parity states show a stronger in-medium modification. 
Requiring that $T_{\rm fo}^{\rm STAR} = T_{\rm fo}^{\rm LHC}$, we find $\kappa_\sigma=0.214\,\pm\,0.045$, where the estimate for the uncertainty in $\kappa_\sigma$ has been obtained by taking the error bars~\cite{Andronic:2017pug} for $T_{\rm fo}^{\rm LHC}$ and $T_{\rm fo}^{\rm STAR}$ into account. This leads to the blue band in Fig.~\ref{fig:ChiralCondensateAtFO2N}. Note in this context that there is an ongoing discussion on the freeze-out temperature estimate for the highest STAR energies regarding feed-down corrections from weak decays which add a further systematic error, see~e.g.~\cite{Proceedings:2017zdf}. 

Iteratively, we can now obtain a new estimate for the freeze-out temperatures based on a dynamical chiral condensate. The result is shown in~\Fig{fig:PDiaChiralFOpointsMod} (blue open squares). As expected, the freeze-out temperature as determined from a dynamical chiral condensate, which couples stronger to the medium at lower collision energies, is smaller than $T_{\rm fo}$ obtained with vacuum hadronic masses (red open circles).

At very large densities or baryon chemical potentials the situation gets increasingly complicated. This applies in particular to the densities achieved at the HADES experiment with $\sqrt{s_{\rm NN}}=2.4$~GeV. In this high density regime interaction terms become more important. It is illustrative to first consider the liquid phase at very low temperatures. Here the chiral condensate drops and we expect a sizable vector condensate $\omega_0\sim n$. The latter leads to an effective reduced baryon chemical potential $\mu_B\to\mu_B+\lambda_\omega n$ with a repulsive vector coupling $\lambda_\omega<0$. At $T=0$, the two effects nearly compensate each other up to a binding energy of $-16$~MeV. Therefore, at very low temperatures the Boltzmann factor is approximately the same as in a gas phase with vacuum mass. Hence, within a linear extrapolation of the gas description into the liquid phase, freeze-out conditions may be obtained from thermal fits using vacuum masses. This argument leads to the triangle for HADES in~\Fig{fig:PDiaChiralFOpointsMod}. 

In the above spirit of a linear extrapolation we also discuss the influence of temperature effects: 
the expectation value of the chiral condensate stays approximately constant for low $T$, while the net-baryon density increases. This leads to an effective increase of the numerator $m_p-\mu_B-\lambda_\omega n$ in the Boltzmann factor. Therefore, in order to reproduce experimentally measured particle multiplicities, the freeze-out temperature at HADES is likely to increase from the current estimate~\cite{HADESFO1,HADESFO2}. This is indicated by the upward arrow emanating from the triangle in \Fig{fig:PDiaChiralFOpointsMod}. The given linear estimate of a highly non-linear problem could reconcile the different freeze-out temperatures at HADES determined from strange versus light hadron multiplicities in line with former estimates from studies of event-by-event multiplicity fluctuations~\cite{Alba:2014eba,Bellwied:2018tkc,Bluhm:2018aei}.

Finally, we predict a system-size independent freeze-out temperature for a system-size scan at LHC energies. By choosing different ion species the fireball volume at chemical freeze-out can be modified. With decreasing $V$, we expect that the coupling of the dynamical chiral condensate to the thermal medium is gradually enhanced, cf.~(\ref{eq:partialtbarsigma}). At lower collision energies indications for an increase of the freeze-out temperatures with decreasing system size have been found from thermal fits using vacuum masses~\cite{Becattini:2005xt}. This is in line with our expectations: taking the chiral dynamics and, thus, an increasing importance of thermal effects on the hadronic masses into account, the freeze-out temperatures for different system sizes should rather coincide, of course, under the assumption that the smaller systems are still large enough to form a thermal medium. 

\section{Conclusions}

We argue that the freeze-out curve in heavy-ion collisions should be located via thermal model fits using dynamical hadronic masses. This dynamics is that of the chiral condensate near the chemical freeze-out and is volume- and density-dependent. Investigating the general form of the evolution equations for the chiral condensate and its fluctuations we find that at high collision energies the chiral condensate decouples from the medium evolution. Instead, it relaxes exponentially fast to its vacuum expectation value. This scenario provides a justification for using vacuum masses in the thermal description of the fireball chemistry at LHC energies.

At lower collision energies the evolution of the chiral condensate couples with increasing strength to the thermal medium. We expect that the impact of the related thermal hadron masses becomes already important for the highest STAR collision energies. Estimating the effect based on the proton multiplicities, we find systematically reduced freeze-out temperatures compared to thermal fits using vacuum masses. At the HADES collision energy standard thermal model fits only work sufficiently well if strangeness is excluded~\cite{HADESFO1}. For these low energies our arguments are only qualitative but potentially allow an explanation of all measured particle yields. 

Finally, we propose a system-size scan at LHC energies: a thermal fit with vacuum masses gives increasing freeze-out temperatures with decreasing system size. According to our scenario the chiral dynamics with its volume-dependent in-medium effects leads to system-size independent freeze-out temperatures. 

\section*{Acknowledgments}

We thank A.~Andronic, M.~Lorenz, C.~Wetterich and T.~Sch\"afer for discussions. MB and MN acknowledge the support by the ExtreMe Matter Institute EMMI, where important contributions to this paper were worked out during their stays as "EMMI Visiting Professors" in Heidelberg. MB and MN further acknowledge the support by the program "Etoiles montantes en Pays de la Loire 2017". 
This work is also supported by the BMBF grant 05P18VHFCA. It is part of and supported by the DFG Collaborative Research Centre SFB 1225 (ISOQUANT) and by the DFG under Germany's Excellence Strategy EXC - 2181/1 - 390900948 (the Heidelberg Excellence Cluster STRUCTURES). 

\bibliography{bib-transport}

\end{document}